\documentclass[twocolumn]{aastex62}
\usepackage{graphicx, amsmath, amssymb, color}
\begin{document}
\title{Band-limited Features in the Primordial Power Spectrum Do Not Resolve the Hubble Tension}
\author{MiaoXin Liu}
\affiliation{School of Physics and Astronomy, Sun Yat-Sen University, 2 Daxue Road, Tangjia, Zhuhai, 519082, P.R.China}
\author{Zhiqi Huang}
\email{huangzhq25@mail.sysu.edu.cn}
\affiliation{School of Physics and Astronomy, Sun Yat-Sen University, 2 Daxue Road, Tangjia, Zhuhai, 519082, P.R.China}

\correspondingauthor{Zhiqi Huang}

\date{\today}
\begin{abstract} 
  For a standard $\Lambda$CDM universe with a power-law primordial power spectrum, the discrepancy between early- and late-universe measurements of the Hubble constant continued to grow and recently reached $5.3\sigma$. During inflation, local features in the inflationary potential often lead to band-limited features in the primordial power spectrum, hence breaking the power-law assumption in the derivation of the Hubble tension. We investigate whether such inflationary ``glitches'' can ease the Hubble tension. The recently released \emph{Planck} temperature and polarization data and the 2019 SH0ES+H0LiCOW joint constraint on the Hubble constant are combined to drive a blind Daubechies wavelet signal search in the primordial power spectrum, up to a resolution $\Delta \ln k\sim 0.1$. We find no significant detection of any features beyond power law. With 64 more degrees of freedom injected in the primordial power spectrum, the Hubble tension persists at a $4.9\sigma$ level.
\end{abstract}


\section{Introduction}\label{sec:intro}

The $\Lambda$ cold dark matter ($\Lambda$CDM) model has been taken as the standard cosmological paradigm since the discovery of late-universe acceleration~\citep{Riess98, Perlmutter99}. It is a remarkable success in terms of explaining the temperature and polarization anisotropies of the cosmic microwave background (CMB) that have been accurately measured by the \emph{Planck} satellite~\citep{Planck2018Overview, Planck2018Params}, the baryon acoustic oscillation features in the galaxy redshift survey data~\citep{BAO-SDSS-DR12-LOWZ, DES1yr_BAO}, the weak gravitational lensing of galaxies~\citep{DES1yr_WL}, the Type Ia supernovae luminosity distances~\citep{Pantheon, Macaulay_2019}, and many others.

Recently, the local distance-ladder measurement of Hubble constant (SH0ES) ~\citep{Riess16, Riess18a, Riess18b, Riess19}, followed by independent support from time delay of strong-lensing quasars images (H0LiCow, STRIDES)~\citep{H0LiCow,STRIDES2019}, starts to challenge the ``concordance'' $\Lambda$CDM picture. Assuming a minimal six-parameter $\Lambda$CDM model, SH0ES and H0LiCOW results together provide a 5.3$\sigma$ difference of $H_0$ with the CMB measurement. This inconsistency, often referred to as ``Hubble tension'', may indicate new physics beyond $\Lambda$CDM. Simple one-parameter extensions of $\Lambda$CDM, however, were found insufficient to resolve the Hubble tension~\citep{Guo_2019, Miao_2018}. More sophisticated models are hence proposed to take the challenge. The list includes but is not limited to modified gravity~\citep{Lin_2018, Lin_2019, Sola_2019,Rossi:2019lgt}, dark energy with phantom equation of state~\citep{Sha2,Sha3,Panpanich19}, early dark energy models~\citep{Karwal_2016,Alexander_2019,Poulin_2019}, backreaction phenomenons~\citep{Racz_2017, Kovacs_2020}, interacting dark components~\citep{DV_2017,Yang_2018a,Yang_2018b,DV_2018,Bhattacharyya_2019, DV_2019}, decaying dark matter~\citep{Vattis_2019, Blinov_2020}, modified recombination history~\citep{Chiang_2018, Gen_2020, LHL_2020},  primordial magnetic fields~\citep{Jedamzik_2020}, and extra relativistic species~\citep{D_Eramo_2018, Benetti_2017, Benetti_2018,Graef_2019,Carneiro_2019}. It has also been claimed that the Hubble tension may just be a relativistic non-linear effect in the standard $\Lambda$CDM paradigm~\citep{Bolejko_2018}. 

Adhikari and Huterer proposed that non-Gaussian CMB covariance from a strong coupling between long-wavelength modes and short-wavelength modes can resolve the Hubble tension~\citep{SuperCMB}. We repeated their calculation and found the same results. However, we noticed that in this model the posterior amplitude of matter fluctuations ($\sigma_8$) is significantly higher than $\Lambda$CDM value,  which is already at the upper edge of the bounds from late-universe observations of galaxy clustering and weak gravitational lensing~\citep{Planck2018Params}. Moreover, it is yet to be shown that the prediction of polarization and the large tri-spectrum in this model is consistent with \emph{Planck} data.

Nevertheless, the idea that Hubble tension may be due to some anomalies in primordial conditions is worth further investigation.

In the concordance picture, the initial seeds of cosmological fluctuations are assumed to originate from vacuum quantum fluctuations during early-universe inflation. For simplest single-field slow-roll inflation models, the predicted primordial metric fluctuations are almost perfectly Gaussian, and has a slightly tilted power-law primordial scalar power spectrum $\mathcal{P}(k)=A_s\left(\frac{k}{k_{\rm pivot}}\right)^{n_s-1}$, where $k$ is the comoving wave number and $k_{\rm pivot}=0.05\mathrm{Mpc}^{-1}$ is the pivot scale. The standard analysis of CMB and large-scale structure data is usually established on this featureless power-law primordial power spectrum. The global deviation from power-law shape, is bounded by \emph{Planck} data within a sub-percent level: $\frac{dn_s}{d\ln k}=-0.0041\pm 0.0067$~\citep{Planck2018Params}, which is fully consistent with the single-field slow-roll prediction $\left\vert\frac{dn_s}{d\ln k}\right\vert \lesssim 10^{-3}$. The \emph{Planck} collaboration also studied a broad class of inflation models as well as many phenomenological parametrizations, but found no evidence beyond the single-field slow-roll scenario~\citep{Planck2018Inflation,Planck2018NG}. Neither does a blind node expansion with cubic-spline interpolation favor any smooth non-power-law features with a resolution $\Delta \ln k\sim 1$~\citep{Planck2018Inflation}. These results are supported by many other independent works~\citep{Meerburg_2012, Zeng_2019, Domenech_2019}. In summary,  the CMB data do not favor any global periodic oscillations or any broad smooth features with resolution $\Delta \ln k \sim 1$.

The apparently missing ingredient - sharper local features with $\Delta \ln k \ll 1$ are as well motivated from the theoretical perspectives. Note that $\ln k$ roughly corresponds to physical time or number of expansion e-folds during inflation. Many slow-roll-breaking processes during inflation, such as crossing a step in the inflaton potential, has strong impact only for $\sim \text{a few} \times 0.1$ efolds. These models can then produce sharp ($\Delta \ln k \sim \text{a few}\times 0.1$) features that are typically local in time ($\ln k$) domain and band-limited in frequency (Fourier conjugate of $\ln k$) domain. One way to study these sharp features is the top-down approach, that is, to parameterize and constrain the predicted features, in a model-by-model manner. For a few templates from popular models, the \emph{Planck} collaboration, again, found null results~\citep{Planck2018Inflation}. See also Refs.~\citep{Hazra2013Reconstruction,Verde2008On,Tocchinivalentini2006Non,handley2019bayesian} for earlier works. The other way, which is missing for the latest Hubble-tension related data, and will be done in this work, is the bottom-up approach that model-independently covers a much broader class of models. 

We apply a wavelet analysis, a statistical tool specifically designed to study local and band-limited signals, to search for sharp features in the primordial power spectrum. Similar analysis has been done for earlier CMB data from COBE and WMAP satellites~\citep{pando1998evidence,mukherjee2000do,mukherjee2003wavelet,mukherjee2003direct,shafieloo2007features}, before \emph{Planck} data drove the Hubble tension. The purpose of our re-examination in the latest \emph{Planck} data is to investigate whether the Hubble tension is driven by a primordial sharp feature that manifests itself in high-$\ell$ multipoles that are only accurately measured by \emph{Planck}.

This paper is organized as follows. In Sec.~\ref{sec:wavelet} we introduce Daubechies wavelet analysis and our power spectrum reconstruction method. In Sec.~\ref{sec:test} we test the wavelet reconstruction method with mock CMB data for a fiducial inflationary model with slow-roll violation. In Sec.~\ref{sec:results} we report the results for \emph{Planck} + SH0ES + H0LiCow data. Sec.~\ref{sec:conclu} concludes.

\section{Method \label{sec:wavelet}}

\begin{center}
\begin{figure}
  \includegraphics[width=0.48\textwidth]{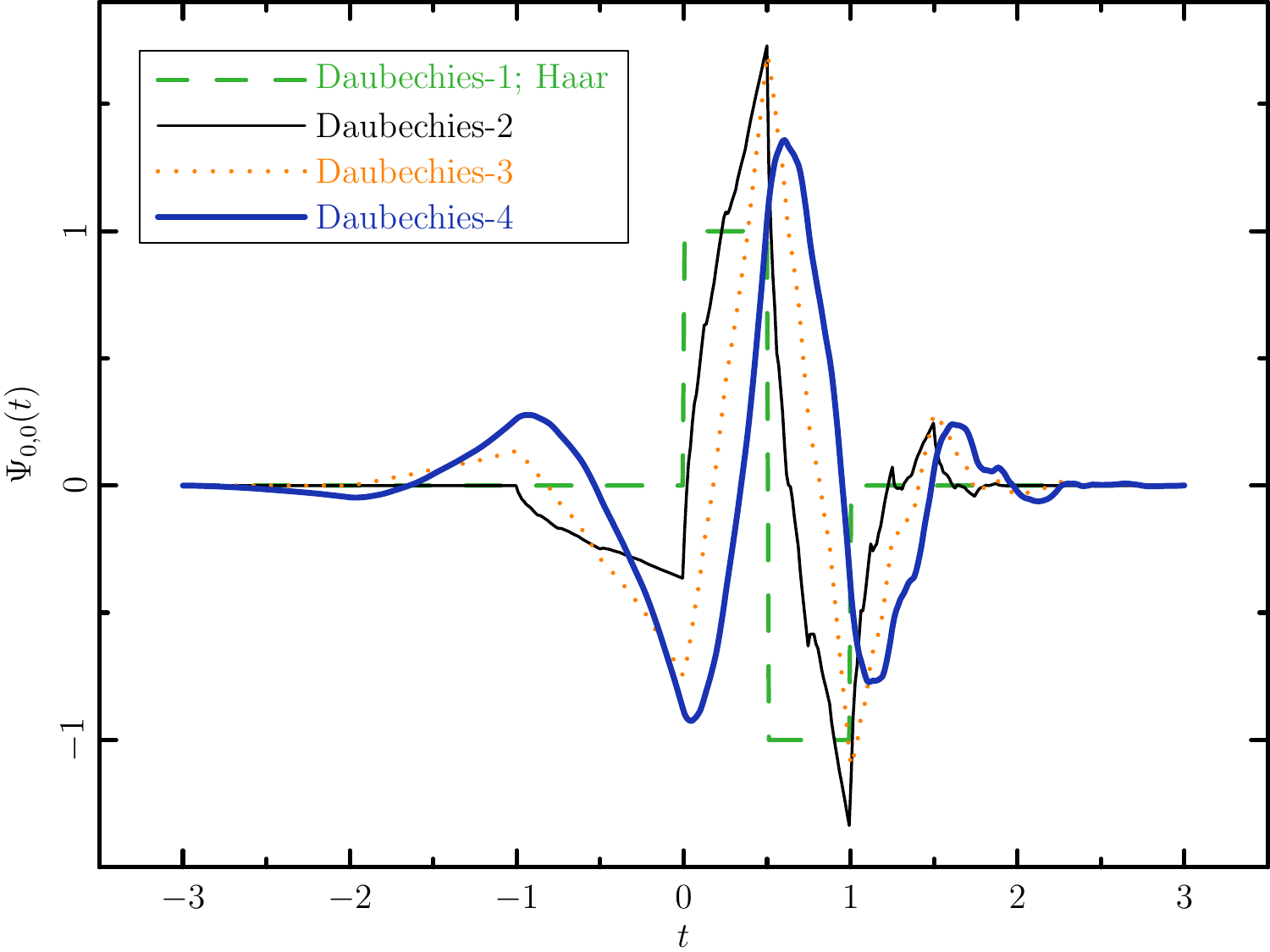}
  \caption{Daubechies wavelet mother functions.\label{fig:wavelets}}
\end{figure}
\end{center}

The Daubechies wavelet basis takes the form:
\begin{equation}
  \Psi_{n, m} (t) = 2^{n/2}\Psi_{0, 0}(2^nt-m), \ \ n,m \in Z
\end{equation}
where $\Psi_{0, 0}$ is the mother function of Daubechies wavelet. The basis functions are complete, compactly supported and orthogonal with respect to both the scale $n$ and the position $m$ indices. They are moving kernels with hierarchical resolutions, with each resolution level a factor of $2$ finer than the previous one. As shown in Fig.~\ref{fig:wavelets}, the Daubechies mother functions are not unique. The most oft-used 1st order Daubechies mother function, also known as Haar wavelet, is simple but discontinuous. Higher-order Daubechies in general cannot be expressed with elementary functions, but are continuous. The smoothness of the Daubechies mother function increases with its order. In this work, we use the 4ts order Daubechies basis, and check the robustness of our result with second-order Daubechies basis. The algorithm to construct Daubechies mother function of arbitrary order is given in Appendix~\ref{append}.

To blindly search features in the primordial scalar power spectrum $\mathcal{P}(k)$, we decompose its deviation from power-law shape into Daubechies wavelets
\begin{equation}
  \ln \frac{\mathcal{P}(k)}{\mathcal{P}_{\rm ref}(k)}  =  \sum_{n=0}^3 \sum_{m=-2^{n+1}}^{2^{n+1}} A_{n,m}\Psi_{n, m} \left(\ln \frac {k}{k_{\rm pivot}}\right), \label{eq:pk}
\end{equation}
where  the reference power-law is $\mathcal{P}_{\rm ref}(k) = A_s\left(\frac{k}{k_{\rm pivot}}\right)^{n_s-1}$. The lower and upper bounds of the scale index $n$ are chosen such that the resolution in $\ln k$ is limited to $ 0.1 \lesssim \Delta\ln k \lesssim 1$, to match features from slow-roll-breaking processes during inflation. The lower and upper bounds of the position index $m$ are chosen such that CMB scales measured by \emph{Planck} are well covered.

We use the  publicly available software CosmoMC~\citep{Lewis2002Cosmological} to run Markov Chain Monte Carlo (MCMC) simulations and to estimate the marginalized bounds of cosmological parameters, which include the standard six built-in parameters $\Omega_bh^2$, $\Omega_ch^2$, $\theta$, $\tau_{\rm re}$, $\ln\left(10^{10}A_s\right)$, $n_s$ and the sixty-four $A_{n,m}$ coefficients defined in Eq.~\eqref{eq:pk}. Here $\Omega_bh^2$ and $\Omega_ch^2$ are baryon and CDM densities, respectively; $\theta$ is the angular extension of sound horizon on the last scattering surface; $\tau_{\rm re}$ is the reionization optical depth; $A_s$ and $n_s$ are the amplitude and index of the primordial scalar power spectrum. The Hubble constant $H_0$ can be derived from these parameters. The sum of neutrino masses is fixed to $\sum m_\nu=0.06\mathrm{eV}$, the minimum value allowed in normal hierarchy picture. Flat priors are applied to all the parameters including the $A_{n,m}$ coefficients. 

The advantage of using wavelets is that they are local by construction. The additional degrees of freedom, despite being many, are not strongly correlated. This significantly accelerates the convergence of MCMC sampling.

\section{Test with Mock Data \label{sec:test}}

To test the viability of the wavelet reconstruction method, we consider a toy model with inflationary potential
\begin{equation}
  V = \frac{3}{4}m^2M_p^2\left(1-e^{-\sqrt{\frac{2}{3}}\frac{\phi}{M_p}}\right)^2\left(1+\epsilon e^{-\frac{\left(\phi-\phi_0\right)^2}{2\mu^2}}\right), \label{eq:infmodel}
\end{equation}
where $M_p=2.45\times 10^{18}\mathrm{GeV}$ is the reduced Planck mass. This potential is constructed by adding a small bump, characterized by the amplitude parameter $\epsilon\ll 1$, the position parameter $\phi_0$ and the width parameter $\mu$, to the Starobinsky potential~\citep{Starobinsky_1983}. The parameters $m = 1.191\times10^{-5}M_p$, $\phi_0 = 5.37 M_p$, $\mu = 0.005M_p$, and $\epsilon = 10^{-5}$ are chosen such that, when instant reheating is assumed, the primordial power spectrum roughly matches CMB observations. The small bump leads to a temporary slow-roll violation and a typical width $\Delta\ln k \sim \text{a few}\times 0.1$ of the feature in the primordial power spectrum, which we compute by numerically integrating the linear perturbation equations of the gauge-invariant Sasaki-Mukhanov variable~\citep{Sasaki_1986, Mukhanov_1988}. The other cosmological parameters for the fiducial cosmology are taken to be the \emph{Planck} 2018 best-fit values~\citep{Planck2018Params}, as shown in the first column of Table~\ref{tab:test}.

To generate the mock CMB data, we assume the full width at half maximum (FWHM) $ = 5\,\mathrm{arcmin}$ for the temperature beam resolution, and $\mathrm{FWHM}=10\,\mathrm{arcmin}$ for polarization. With Gaussian approximation, the mock CMB likelihood reads~\citep{Verde/etal:2006}
\begin{equation}
\begin{split}
\label{chisq_CMB}
\ln\mathcal{L} =& -\frac{f_{\rm sky, eff}}{2}\sum_{\ell=\ell_{\min}}^{\ell_{\max}} (2\ell+1) \\
  &\times \left[\frac{\hat{{\cal C}}_\ell^{TT}{\cal C}_\ell^{EE} + \hat{{\cal C}}_\ell^{EE}{\cal C}_\ell^{TT} - 2\hat{{\cal C}}_\ell^{TE}{\cal C}_\ell^{TE}}{{\cal C}_\ell^{TT}{\cal C}_\ell^{EE}-({\cal C}_\ell^{TE})^2} \right. \\ 
& + \left. \ln{\left(\frac{{\cal C}_\ell^{TT}{\cal C}_\ell^{EE}-({\cal C}_\ell^{TE})^2}{\hat{{\cal C}}_\ell^{TT}\hat{{\cal C}}_\ell^{EE}-(\hat{{\cal C}}_\ell^{TE})^2}\right)} - 2\right] \ ,  
\end{split}
\end{equation}
where we have used $\ell_{\min}=2$, $\ell_{\max}=2500$, and an effective sky coverage $f_{\rm sky,eff} = 0.85$. In this formula, ${\cal C}^{XY}_\ell$  ($X,Y \in \{T, E\}$) are the model-dependent theoretical angular power spectra. They are given by ${\cal C}^{XY}_\ell= C^{XY}_\ell + N_\ell^{XY}$, where $C_\ell^{XY}$ are the noise-free CMB power spectra calculated with the publicly available code CAMB \citep{Lewis/etal:2000} and $N_\ell^{XY}$ are the noise spectra. To simulate the noise spectra, we assume a Gaussian beam shape and a sensitivity $50\,\mathrm{\mu K\,s^{1/2}}$ for temperature and $100\,\mathrm{\mu K\,s^{1/2}}$ for polarization, both integrated for five years. The hatted symbols $\hat{\cal C}_\ell^{XY}$ represent the mock data predicted by the fiducial cosmology. To check whether the reconstruction method produces any bias, we do not add a realization of cosmic variance onto the mock data. Thus, any significant deviation from the fiducial model should be interpreted as a bias rather than a look-elsewhere effect. For the real data that we will discuss in the next section, the look-elsewhere effect cannot be avoided and weak ``anomalies'' should not be overly interpreted.

\begin{figure}
  \includegraphics[width=0.48\textwidth]{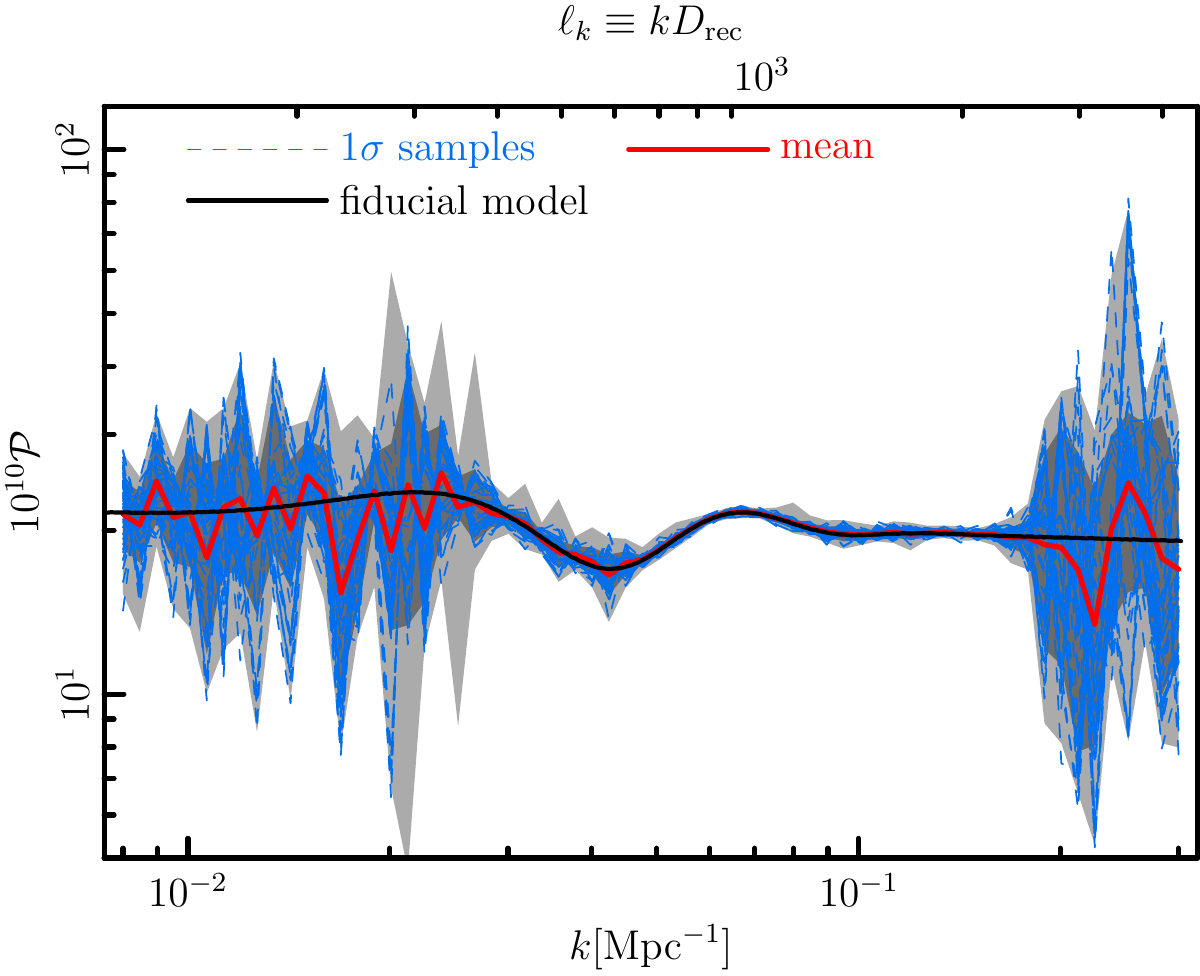}
  \caption{Reconstructed primordial power spectrum for the mock CMB data generated from the inflationary model in Eq.~\eqref{eq:infmodel}. The dashed sky-blue lines are randomly picked trajectories from the likelihood-ordered top 68.3\% MCMC samples. The dark-gray and light-gray contours are  marginalized 68.3\% and 95.4\% confidence level bounds, respectively.\label{fig:test}}
\end{figure}

We apply the wavelet reconstruction method to the mock CMB data. An unbiased detection of the input power spectrum including the slow-roll violation signal is shown in Fig.~\ref{fig:test}. As shown Table~\ref{tab:test}, the other input cosmological parameters are also well recovered with no noticeable bias.

\begin{table}
  \begin{center} 
    \caption{Marginalized constraints on cosmological parameters for mock data.}
    \label{tab:test}
    \begin{tabular}{ccc}
      \hline
      \hline
      Parameter & fiducial & constraint \\
      $\Omega_b h^2$ & $0.02238$ & $0.02233^{+0.00036}_{-0.00031}$ \\ 
      \hline
      $\Omega_c h^2$ & $0.1201$ & $0.1202^{+0.0026}_{-0.0025}$ \\ 
      \hline
      $H_0$ (km/s/Mpc) &  $67.3$ & $67.2^{+1.0}_{-1.0}$ \\
      \hline
      $\tau_{\rm re}$ & $0.0543$ & $0.0542^{+0.0027}_{-0.0026}$ \\ 
    \hline
    \end{tabular}
  \end{center}
\end{table}

We leave more detailed interpretation of the reconstructed primordial power spectrum to the next section, where the real CMB data are investigated. 

\section{\emph{Planck} + SH0ES + H0LiCow \label{sec:results}}

To explicitly extract Hubble-tension-driven wavelet signals, we use jointly the SH0ES + H0LiCow constraint $H_0 = 73.82\pm 1.10 \,\mathrm{km\,s^{-1}Mpc^{-1}}$ ~\citep{H0LiCow} and the \emph{Planck} final release of TT,TE,EE + lensing likelihood~\citep{Planck2018Like}. Unlike the idealized mock CMB data that we discussed in the last section, the \emph{Planck} likelihood contains many nuisance parameters to describe uncertainties in the foreground template, etc., all of which are marginalized over in our analysis.

\begin{table}
  \begin{center} 
    \caption{Marginalized constraints on cosmological parameters for \emph{Planck}+SH0ES+H0LiCow.}
    \label{tbl:cosmomc}
    \begin{tabular}{cc}
      \hline
      \hline
      $\Omega_b h^2$ & $0.0232^{+0.0005}_{-0.0005}$ \\ 
      \hline
      $\Omega_c h^2$ & $0.1169^{+0.0015}_{-0.0016}$ \\ 
      \hline
      $100\theta_{MC}$ & $1.04146^{+0.00037}_{-0.00037}$ \\ 
      \hline
      $\tau_{\rm re}$ & $0.063^{+0.010}_{-0.008}$ \\ 
      \hline
      ${\rm{ln}}(10^{10} A_s)$ & $3.060^{+0.019}_{-0.018}$ \\ 
      \hline
      $n_s$ & $0.995^{+0.010}_{-0.010}$ \\
      \hline
      $H_0$ (km/s/Mpc) &  $69.4^{+0.7}_{-0.7}$\\ 
      \hline
      $A_{3,-7}$ & $-0.033^{+0.015}_{-0.015}$ \\
      \hline
      other $A_{n,m}$'s & no detection beyond $2\sigma$ \\
      \hline
    \end{tabular}
  \end{center}
\end{table}

Table~\ref{tbl:cosmomc} lists the marginalized $1\sigma$ constraints of cosmological parameters. The sixty-four wavelet expansion coefficients are mostly consistent with zero within $2\sigma$, with only one $2.2\sigma$ exception $A_{3, -7}=-0.033\pm 0.015$. The $2.2\sigma$ weak anomaly can be well explained by look-elsewhere effect for the many degrees of freedom we have injected into the model. Another week anomaly is in the posterior of the reference $n_s=0.995\pm 0.010$, which is $\sim 2.7\sigma$ higher than the ``no wavelet, no $H_0$ prior'' case $n_s=0.965\pm 0.005$~\citep{Planck2018Params}. This can be explained by the known positive correlation between $n_s$ and $H_0$. More interestingly, it has been shown that the combination \emph{Planck} + SH0ES favors a model with a scale invariant primordial power spectrum and $\sim 0.7\pm 0.13$ extra relativistic species~\citep{Benetti_2017, Benetti_2018}. 

Finally, we would like to point out that these weak anomalies are not associated with the wavelet reconstruction method or our particular choice of the wavelet mother function, as the anomalies do not show up in the test with mock CMB data where the look-elsewhere effect is avoided on purpose. 

\begin{center}
\begin{figure}
  \includegraphics[width=0.48\textwidth]{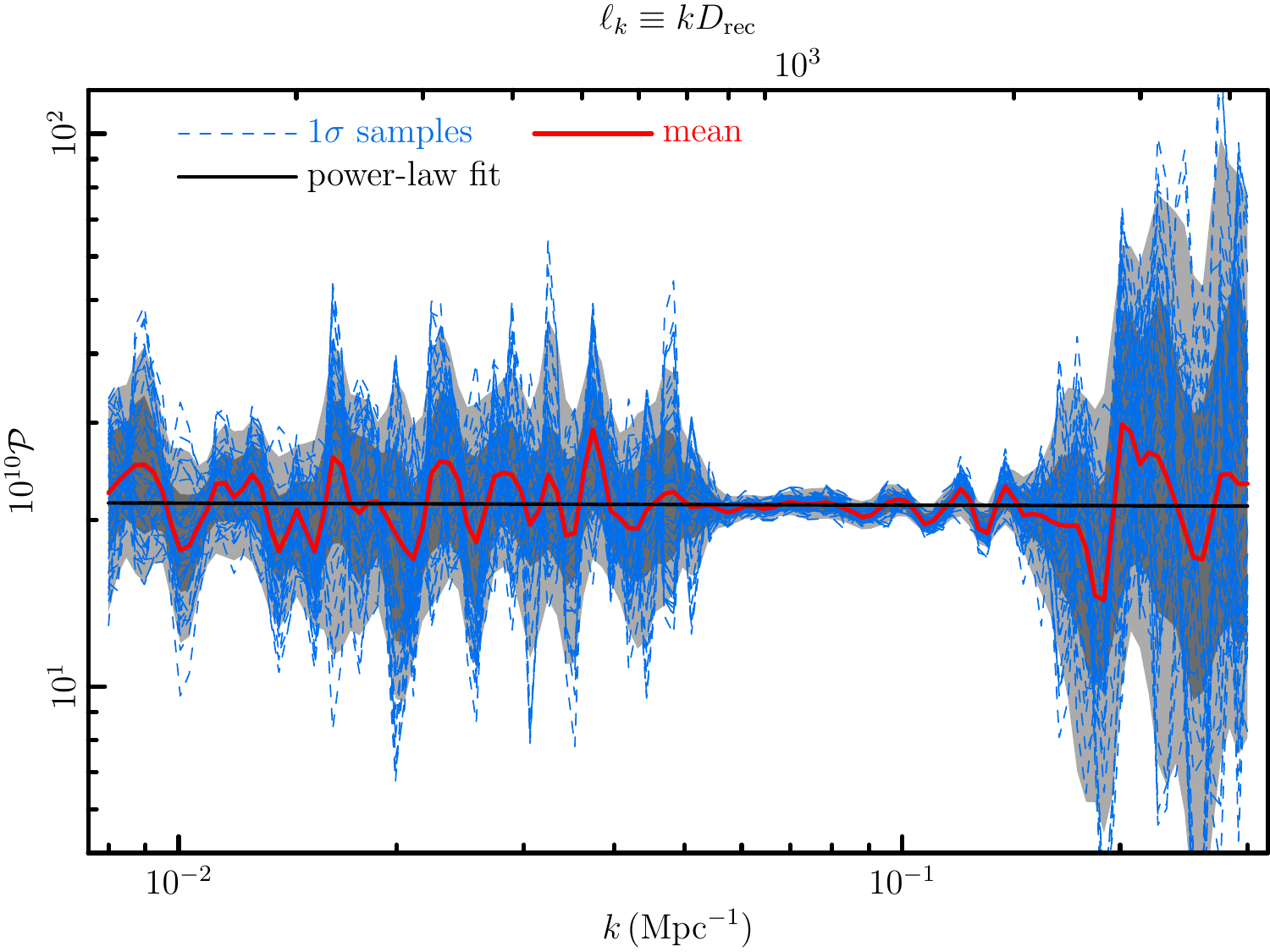}
  \caption{Reconstructed primordial power spectrum for \emph{Planck}+SH0ES+H0LiCow. The dashed sky-blue lines are randomly picked trajectories from the likelihood-ordered top 68.3\% MCMC samples. The dark-gray and light-gray contours are  marginalized 68.3\% and 95.4\% confidence level bounds, respectively.\label{fig:trajs}}
\end{figure}
\end{center}

In Fig.~\ref{fig:trajs} we again visualize the reconstructed $\mathcal{P}(k)$ trajectories. The non-deviation from a power-law spectrum is consistent with the posteriors of the $A_{mn}$ parameters. The constraints are worse than the test case with mock CMB data, because the real \emph{Planck} data has foreground modeling uncertainties (especially for the polarization) and a slightly higher noise level than what we assumed in Sec.~\ref{sec:test}.

\begin{center}
\begin{figure}
  \includegraphics[width=0.48\textwidth]{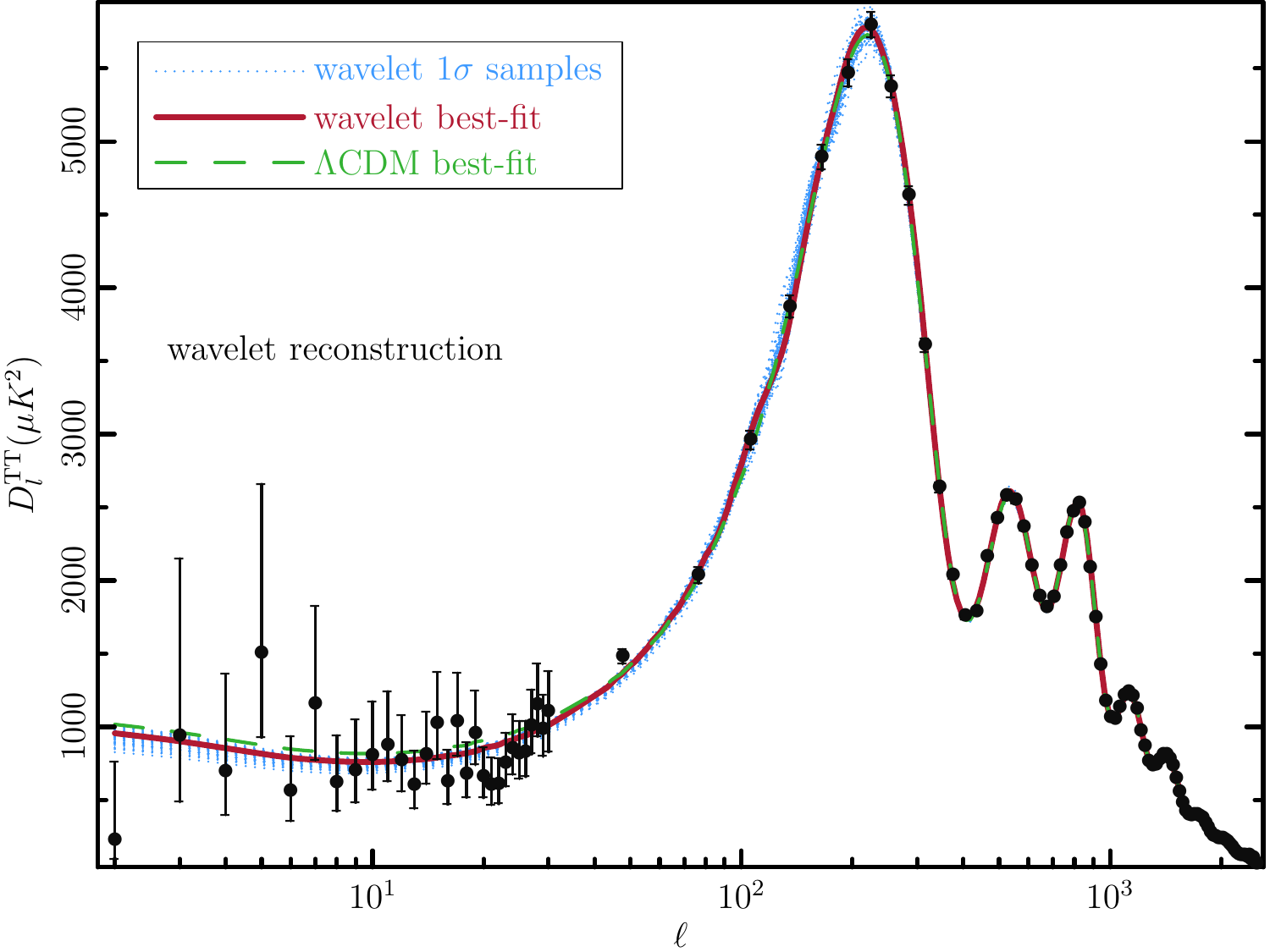}
  \caption{CMB temperature power $D_\ell^{\rm TT}\equiv \frac{\ell(\ell+1)}{2\pi}C_\ell^{\rm TT}$, where $C_\ell^{\rm TT}$ is the angular power spectrum of temperature fluctuations. The dotted sky-blue lines are randomly picked trajectories from the likelihood-ordered top 68.3\% MCMC samples. The solid red line is the best-fit for wavelet expansion of $\mathcal{P}(k)$, whereas the dashed green line is the best-fit for the minimal six-parameter $\Lambda$CDM with power-law $\mathcal{P}(k)$.\label{fig:TT}}
\end{figure}
\end{center}

Compared to the 12-knot cubic spline reconstruction in section 6.3 of \citet{Planck2018Inflation}, our wavelet analysis, by construction, picks out more local and sharper features. The high-frequency wiggling in $\mathcal{P}(k)$ is driven, or at least partially driven by the statistical fluctuations in CMB power spectra. In Fig~\ref{fig:TT} we show how the wavelet trajectories follow statistical fluctuations in the temperature angular power spectrum $D_\ell^{\rm TT}$, allowed by cosmic variance at low and intermediate ell's. At higher ell's, the trajectories converge due to a much smaller cosmic variance. These features can also be seen in the left and middle parts of Fig~\ref{fig:trajs}. The large scattering in the right part of Fig~\ref{fig:trajs} corresponds to the unconstrained power on small scales  (high-$k$)  beyond \emph{Planck} resolution.

\begin{center}
\begin{figure}
  \includegraphics[width=0.48\textwidth]{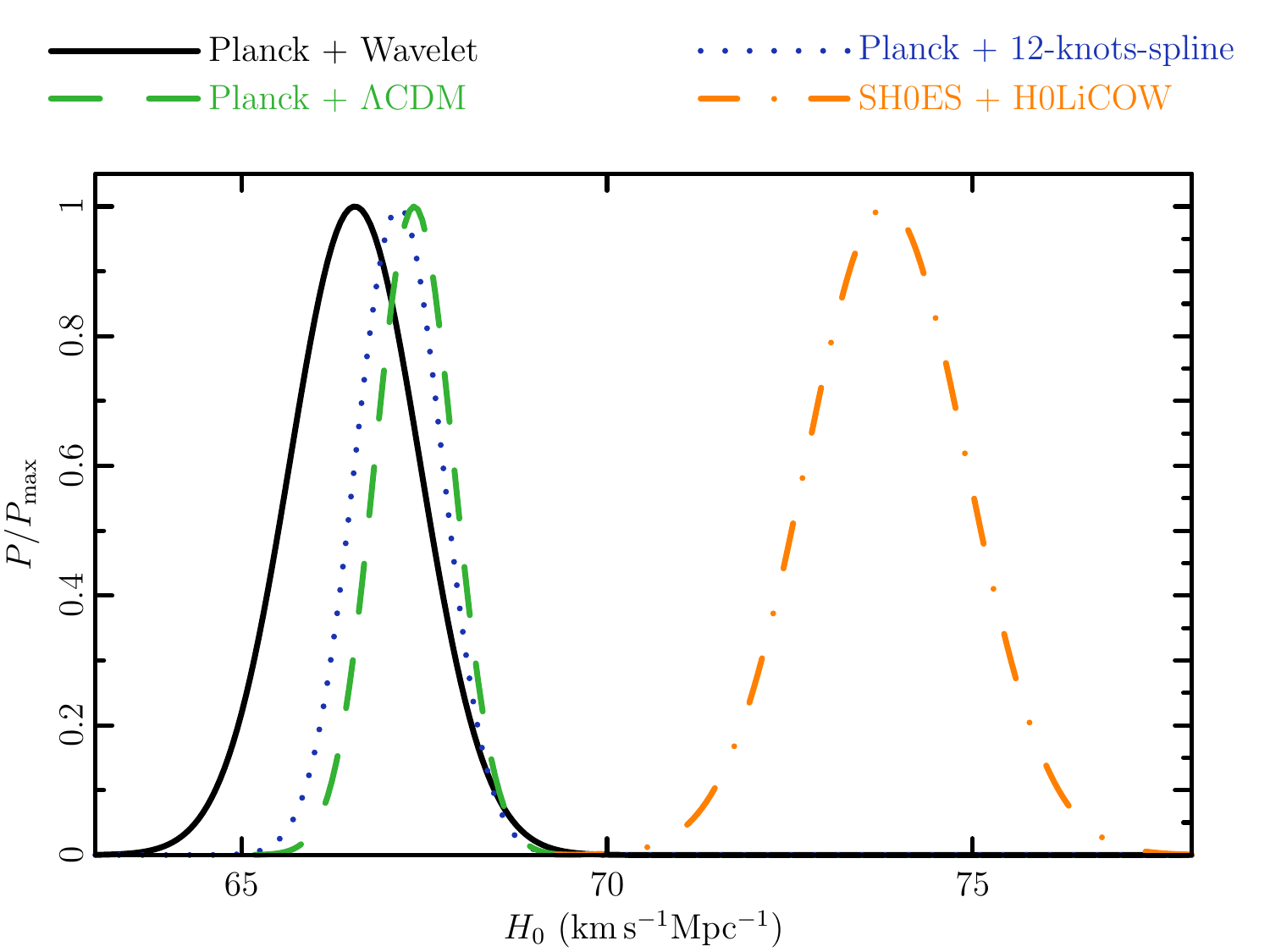}
  \caption{Comparison of $H_0$ constraints.\label{fig:H0}}
\end{figure}
\end{center}

For the Hubble constant, we obtain a \emph{Planck} + SH0ES + H0LiCow joint constraint: $H_0=69.4\pm 0.7\mathrm{km\,s^{-1}Mpc^{-1}}$. Because the posterior is very close to Gaussian, we can approximately remove the SH0ES + H0LiCow contribution and obtain a \emph{Planck}-only constraint, as shown in Fig.~\ref{fig:H0}. For a comparison, we also plot the \emph{Planck} constraint for the standard $\Lambda$CDM power-law case as well as the 12-knot-spline case, which we obtain by repeating the calculations in~\citet{Planck2018Inflation}. We find that allowing more features in the primordial power spectrum, either local and band-limited as in the wavelet case, or just low-pass filtered as in the 12-knot-spline case, in general pushes the mean $H_0$ towards an even smaller value, which balances out the increased uncertainty and keeps the Hubble tension at roughly the same level. More specifically, the tension between \emph{Planck} and SH0ES + H0LiCow is $4.9\sigma$ for the wavelet analysis, and $5.3\sigma$ for the 12-knot-spline.

\section{Conclusion and Discussion \label{sec:conclu}}

To check the robustness of the wavelet analysis method, we repeated our calculation with 2nd order Daubechies basis, and found no significant variations in the results. We thus conclude that the Hubble tension cannot be eased by band-limited features in the primordial power spectrum with $\Delta\ln k\gtrsim 0.1$. 

Ideally, the wavelet analysis, if expanded to infinite order, is equivalent to many other binning, expansion and interpolation methods. Practically, however, one has to introduce a cut in expansion order or a smoothing scheme to reduce the dimension of parameter space and to achieve MCMC convergence. The cut or smoothing schemes in different methods introduce model-dependent priors. In our wavelet analysis, the cut of expansion order leads to a prior that captures the local and band-limited features that naturally arise from various inflationary processes beyond slow-roll. Indeed, if not limited by the physical prior, a deconvolution scheme can map the $H_0$-discordance in the CMB power spectrum to the primordial power spectrum and ease the Hubble tension~\citep{Sha1}.

\section{Acknowledgments}

We thank J. Richard Bond, Lev Kofman and Pascal Vaudrevange for many useful discussions in the memorable days in Toronto. This work is supported by Sun Yat-sen University Research Starting Grant 71000-18841232.

\appendix
\section{The calculation of Daubechies mother wavelet  \label{append}}

The mother wavelet $\Psi_{0,0}(t)$ is constructed from the scaling function $\phi(t)$ (or father wavelet) via
\begin{equation}
\Psi_{0, 0} (t) = \sqrt{2}\sum_{k=0}^{2n-1}(-1)^k c_{2n-1-k} \phi(2t-k), \, t\in(-\infty,\infty),
\end{equation}
where $n$ is the order of the wavelet system. The scaling function $\phi$ is defined recursively through the use of dilation equations. The basic dilation equation reads
\begin{equation}
\phi_m (t) = \sqrt{2}\sum_{k=0}^{2n-1} c_{k} \phi_{m-1}(2t-k) \, .
\end{equation}
It is a two-scale or dyadic difference equation. The starting point of recursion $\phi_{-1}(t)$ is defined as the Haar father wavelet, which is equal to $1$ in the interval $[0,1]$ and vanishes elsewhere. When the recursion order $m \to \infty$, $\phi_m(t)$ converges to the scaling function $\phi (t)$.

The numerical values of the filter coefficients $c_{k}$ can be calculated via the algorithm named after Daubechies ~\citep{Daubechies1998Ten}. Defining the polynomial
\begin{equation}
P_{n} (y) = \sum_{k=0}^{n-1} \binom{n-1+k}{k} y^k\, ,
\end{equation}
one can calculate numerically the (complex) roots of 
\begin{equation}
P_{n} (\frac{1}{2}-\frac{1}{4z}-\frac{z}{4})\, ,
\end{equation}
where $z$ is an arbitrary complex variable. Selecting those roots $r_i$ inside the unit circle ($\|r_i\|<1$), one obtains the filter coefficients by identifying the coefficients $\tilde{c_k}$ of
\begin{equation}
(z+1)^n \prod_{i=1}^{n-1} (z-r_i) \equiv \sum_{k=0}^{2n-1} \tilde{c_k} z^{2n-k-1}\, .
\end{equation}
After normalization, the coefficients are
\begin{equation}
c_k=\frac{\tilde{c_k}}{\sqrt{\sum_{k=0}^{2n-1} \tilde{c_k}^2}}.
\end{equation}
For example, the $c_0\sim c_7$ of the 4th order Daubechies wavelet system are:
$0.230378$, 
$0.714847$, 
$0.630881$,  
$-0.0279838$, 
$-0.1870348$,  
$0.0308414$,  
$0.0328830$,  
$-0.0105974$.


\end{document}